# Determining the Applicability of Agile Practices to Mission and Life-critical Systems

Ahmed Sidky, James Arthur

(asidky@vt.edu, arthur@vt.edu)

*Virginia Tech*

## Abstract

*Adopting agile practices brings about many benefits and improvements to the system being developed. However, in mission and life-critical systems, adopting an inappropriate agile practice has detrimental impacts on the system in various phases of its lifecycle as well as precludes desired qualities from being actualized. This paper presents a three-stage process that provides guidance to organizations on how to identify the agile practices they can benefit from without causing any impact to the mission and life critical system being developed.*

## 1. Introduction

Lately the number of organizations adopting agile practices and concepts is increasing. This increase consists not only of more small teams developing simple applications, but also of large teams successfully developing complex systems [19] [28]. This is a surprise, because initially agile development was considered suitable only for small organizations producing simple applications [13]. However, in spite of this unexpected success, organizations developing mission and life-critical systems are still turning away from fully adopting agile practices for reasons that are both right and wrong.

On the one hand, it is wrong for organizations to reject agile practices all together, because in doing so, they fail to take advantage of the many improvements agile practices can bring to their software development process. Empirical evidence has shown that embracing agile practices yields many benefits [33] [27] [8] [7] [23] [21], including:

- Early return on investment
- Short time to market
- Improved quality
- Enhanced client relationships
- Better team morale.

On the other hand, organizations developing mission and life-critical systems are right to refuse to adopt all of the agile practices and concepts, because the use of some practices can be detrimental to the system during different phases of its lifecycle. Moreover, even though agile development has proved its utility in many domains, many still do not consider it suitable for the development of mission and life-critical systems.[2] [25] [11] [14].

The resolution to this predicament consists of adhering to a process that can provide enough guidance to answer the following question:

*How can an organization identify the agile practices that improve the software development process without causing detrimental impacts to any aspect of mission and life-critical systems?*

Finding the answer to this question presents a challenge to the developers of mission and life-critical systems. The three fundamental factors representing this challenge are:

1. *Identifying the ability of the organization to adopt agile practices*. Regardless of the application, since agile adoption is a form of process improvement and change, we must first verify that the organization is ready to engage in a change towards agility.

2. *Determining the suitability of agile practices in the development of mission and life-critical systems.* What causes certain agile practices to be beneficial (or detrimental) for the development of mission and life-critical systems; and how can developers identify which practices are or are not appropriate?

3. *Determining the suitability of agile practices for the organization developing mission and life-critical systems.* Concluding that a practice is suitable for mission and life-critical systems is one issue. Determining whether the organization possesses the necessary characteristics to adopt that practice successfully is another issue. The former does not imply the latter; and the latter is focus of this challenge.

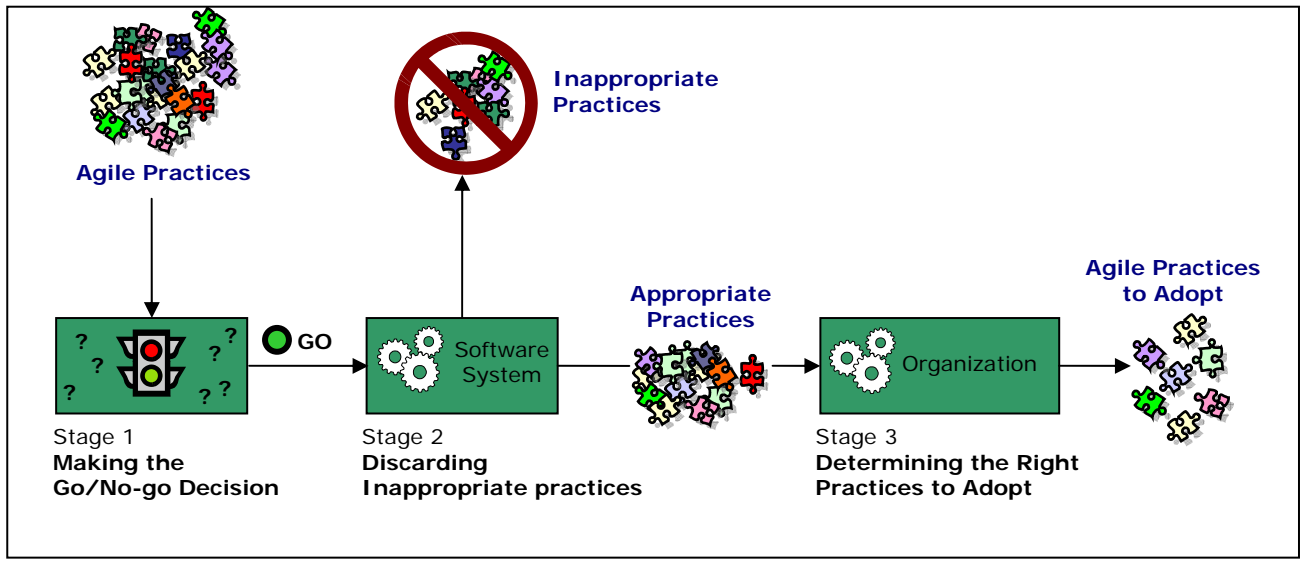

**Figure 1: A 3-Stage Process to determine applicability of individual agile practices**

In this paper, we describe a 3-stage process that provides organizations with guidance on how to determine the applicability of individual agile practices with respect to mission and life-critical systems (see Figure 1). This process and its stages are inspired from the Agile Adoption Framework [29]. These three stages are:

- Stage 1: Making the Go/No-go decision
- Stage 2: Discarding inappropriate practices
- Stage 3: Determining the right practices to adopt

Each stage of the process addresses one of the challenges identified earlier. In this paper, Section 2 describes how the first stage of the framework identifies whether or not the organization is able undergo an agile adoption initiative. Section 3 provides a discussion of how the second stage determines which agile practices are or are not suitable for mission and life-critical systems. Section 4 presents how to assess the organization's readiness to adopt the agile practices best suited for it. Our conclusions are given in Section 5.

## 2. Identifying the Ability of the Organization to Adopt Agile Practices

Identifying whether or not an organization is capable of adopting *any* agile practices is the first challenge that needs to be addressed. Section 2.1 explains the importance of this challenge and the benefits of resolving it. Section 2.2 presents the steps taken in the first stage of the process to determine whether or not the organization is capable of adopting agile practices. The subsequent subsections provide details of these steps.

### 2.1. Importance of Identifying the Organization's Ability to Adopt Agility

In traditional Software Process Improvement (SPI) (based on the CMM and CMMI) it is highly recommended that the decision to start a SPI initiative be taken after a pre-assessment phase is conducted. The pre-assessment determines whether or not the organization is ready for SPI [18]. Organizations that do not embody the factors necessary for a successful SPI effort are considered "not ready." In this situation the SPI effort is suspended until the missing factors become present. This pre-assessment phase is important, because it saves the organization much time, money and effort by identifying upfront that the SPI initiative is doomed to fail. [20]

Since introducing agile practices into a software development process is a type of SPI, a pre-assessment process should be initiated. The successful adoption of agile practices requires the organization to possess certain factors, including the capability of absorbing the cost, as well as expending the required time and effort. Therefore, conducting an initial assessment to determine whether or not the necessary factors for success exist is both beneficial and necessary.

In addition to the above, conducting an assessment in order to decide whether or not to go ahead with the effort to introduce agile practices into mission and life-critical applications is also important because of additional losses that can be incurred through *Technical Chaos*. Technical chaos, the disruptions caused by the partial adoption of new practices, leaves the development process in an unstable state until it reverts back to the original engineering practices used before the failed adoption effort. Technical chaos is likely to occur when an agile adoption effort starts and then fails before completion. Thus, technical chaos can have a drastic effect on mission and life-critical systems as some vital information might be lost during this phase of instability. Therefore, the decision to adopt agile practices for mission and life-critical systems should only be taken after ensuring that the necessary success factors are present in the organization.

The next section describes how Stage 1 of the proposed process guides and assists organizations in making Go/No-go decisions concerning the adoption of agile practices.

### 2.2. Stage 1: Making the Go/No-go decision

The objective of the first stage of the process is to provide organizations with a method for reaching a decision of whether or not to proceed with agile adoption initiatives. Stage 1 suggests the following three steps as a means to fulfill this objective:

1. Identify success factors
2. Assess the extent to which the success factors are present
3. Make Go/No-go decision based on assessment results.

Figure 2 illustrates each of the three steps in Stage 1. A description of each of these steps follows in the next three sub-sections.

**2.2.1. Identify success factors.** The purpose of this step is to determine whether factors are present to support the process improvement effort towards agility, regardless of the individual agile practices the organization chooses to adopt. Success factors can vary from one organization to another. These factors typically pertain to an organization's resources including money, time, and effort, as well as the support of the executives.

For example, if no funds are available to support the agile adoption effort, then the adoption process is not feasible. Therefore, *Availability of Sufficient Funds* is a success factor. *Executive Support* (for the adoption of agile practices) is another success factor. If either of these factors are absent, then the organization is not ready for the adoption process and should not proceed.

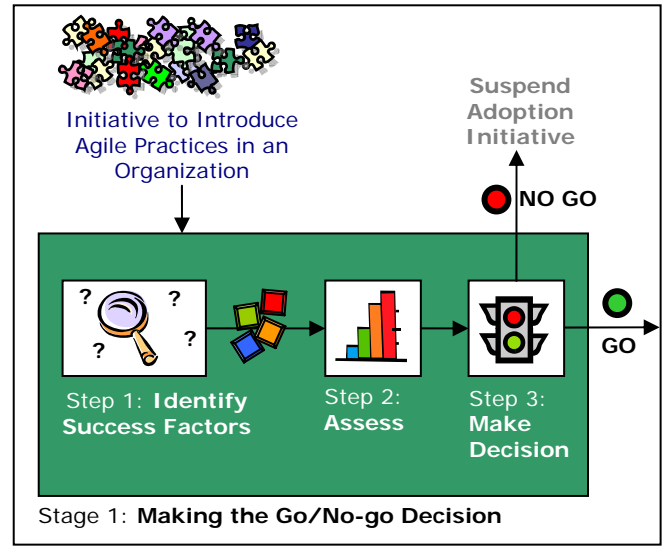

**Figure 2: The three steps of Stage 1**

**2.2.2. Assess the extent to which the success factors are present.** Once the organization identifies the success factors, the next step is to assess the extent to which they are present. Indicators used to assess the degree to which a success factor is present or absent are questions that people in the organization or an assessor answers. The indictors are used to measure the characteristics or aspects needed to identify the presence of a success factor.

For example, one of the characteristics measured to determine the *Availability of Sufficient Funds* is the dollar amount of funds allocated to the process. Another aspect is the ability to actually spend the funds for agile adoption. At least one indicator (more are preferable) is used to assess each aspect or characteristic of a success factor. An example of a question (indicator) used to assess the ability to spend funds on agile adoption; *Can the funds be spent towards any process improvement activity?* Another indicator can be the following: *Are there any restrictions on the type of activities these funds can be used for?*

The answers to all the questions related to the characteristics or aspects of a success factor collectively indicate the extent to which that success factor is present or absent in the organization.

**2.2.3. Make Go/No-go decision based on assessment results.** Once the assessment process has identified the degree to which each success factor is present in the organization, the decision of whether or not to proceed with the agile adoption effort can be made. The "green light" is given when the degree of presence for each of the success factors is above an acceptable threshold for the organization. For example, an organization decides to go ahead when the success factors' degrees of presence all rise above 80%. If one or more of the success factors fall below the threshold, the assessor recommends a *No-go* decision. Clearly, it is necessary to ensure that organizations producing mission and life-critical systems have the capability to introduce agile practices into their process before starting such a process improvement effort. If they are not ready, then they may unnecessarily commit to an initiative, which can have detrimental consequences later.

In summary, Stage 1 provides guidance to organizations needing to decide whether to start the agile adoption effort or not. Identifying and assessing the presence of success factors in the organization determines an organization's ability to proceed with the introduction of agile practices.

## 3. Determining the Suitability of Agile Practices in the Development of Mission and Life-critical Systems

The journey of introducing agile practices in processes for developing mission and life-critical systems begins once the organization concludes that all the success factors needed to support an agile adoption initiative are present. However, for organizations developing mission and life-critical systems, adopting *all* agile practices is not a viable option because some agile practices are simply not suitable. This is the premise behind the second challenge, that is, identifying which agile practices to introduce when developing mission and life-critical systems. Finding an approach for determining the suitability of individual agile practices is needed to address this challenge.

Section 3.1 describes why some agile practices are not suitable for mission and life-critical systems. Section 3.2 presents the second stage of the predictability process that provides an approach for determining the suitability of individual agile practices for mission and life-critical systems.

### 3.1. Why Some Agile Practices are not Suitable for Mission and Life-critical Systems

The investment of time and effort to determine why some agile practices are unsuitable for mission and life-critical systems benefits an organization because it forestalls potential problems. Since system/development characteristics play a large role in determining the suitability of a particular agile practice, it is important to recognize that mission or life-critical systems are usually characterized as being large, complex, and having long development periods. Additional factors affecting the suitability of agile practices relate to the desired system qualities. Such qualities refer to the desired benchmarks that describe the system's intended performance within the environment for which it is built. For example, *Maintainability* and *Reliability* are common desired system qualities for mission and life-critical systems.

If adopting an agile practice conflicts with any system/development characteristic or precludes any desired system quality of mission and life-critical systems, then the practice is considered unsuitable. The following two sub-sections provide examples explaining and demonstrating how two different agile practices are determined unsuitable for mission and life-critical systems.

**3.1.1. Minimal Documentation.** *Minimal Documentation* is a prevalent agile practice. It discourages writing unnecessary requirements, design or management documents for two main reasons [9] [22]: (1) Agile practitioners have recognized that there is a cost associated with developing and maintaining documentation, and (2) documentation does not deliver value to the customer, while working software does. Therefore, minimal documentation is promoted in agile development, because it gives the customer a better and earlier return on investment – valuable working software instead of perceived invaluable documents.

Minimal documentation assumes that enhanced informal communication among team members, known as tacit knowledge, is the primary means of knowledge exchange, not formal documents [7]. As efficient as minimal documentation is as an agile practice, its adoption in mission and life-critical systems is not appropriate. Here are some reasons.

First, mission and life-critical systems typically have long (multiyear) development efforts. Due to the length of the development process, it is highly unlikely that all of the people who start with the project are going to remain until the end – personnel turnover and re-allocations are commonplace. As a result of these circumstances, technical information for training and support needs to be provided by means of documentation that is explicit, comprehensive, and persistent. Since minimal documentation depends on tacit knowledge, using this practice prevents the fulfillment of documentation needs that the extended duration of the development cycle mandates. Therefore, minimal documentation is not a suitable agile practice for mission and life-critical systems [12].

Secondly, minimal documentation is not suitable for mission and life-critical systems because of the detrimental impact the practice has on a highly desirable system quality – *Maintainability*. Since third-party organizations usually handle the maintenance of main mission and life-critical systems, relying on tacit knowledge to communicate and transfer technical information is not a viable option. Minimal documentation will fail to provide the necessary documented information that is needed by third-party organizations in maintaining the system for extended periods of time.

**3.1.2. Evolutionary Requirements.** Another agile practice not suitable for mission and life-critical systems is *Evolutionary Requirements,* a practice that suggests that requirements should evolve over the course of many iterations rather than being fully developed in a major upfront specification effort [22]. Agile practitioners believe that since frequent and constant change should be expected, trying to elicit *all* the requirements upfront is a waste of time, because they will change. Therefore, only eliciting the requirements that are currently known and certain, and leaving the rest to evolve based on the customer's feedback, yields the best return on investment. Despite the suitability of Evolutionary Requirements as an agile practice, it is not suitable for mission and life-critical systems. Some of the reasons behind this inappropriateness are as follows.

Firstly, high reliability is a desired system quality of mission and life-critical systems because it helps promote safety [35]. One of the approaches used to attain high reliability in a system is the use of *Independent Verification and Validation (IV&V)* [4] [17]. One activity within the IV&V of mission and life-critical systems is a *Safety Impact Analysis* that is conducted before the design and coding phases [34]. After verifying that the requirements are complete and comprehensive, the safety impact analysis ensures that they do not jeopardize the safety of the system being developed. Therefore, conducting a safety impact analysis helps to establish the reliability of the system. However, for a safety impact analysis to be conducted, all requirements must be known before any actual development occurs. Adopting evolutionary requirements will fail to meet the needs of the safety impact analysis, hence causing IV&V's to fall short of completing their tasks. Therefore, evolutionary requirements is not a suitable agile practice, because adopting that practice precludes mission and life-critical systems from achieving the highest attainable level of reliability.

Secondly, this practice is not suitable because of a common characteristic found in mission and life-critical systems – software complexity. When adopting the evolutionary requirements approach the design of the system also becomes evolutionary since not all the requirements are known ahead of time. Agile development addresses the issue of continuously developing and improving a system's design by the use of another practice – *Refactoring* [16] [3]. Refactoring involves rewriting the code to improve its structure, while explicitly preserving its behavior. Therefore, regularly refactoring the code is necessary if evolutionary requirements are adopted. In mission and life-critical systems refactoring is a risky and costly task due to size and complexity of the systems [32]. As a result, adopting evolutionary requirements is not suitable for mission in life critical systems because it too, conflicts with a characteristic of the system (high complexity).

These two examples illustrate how the suitability of agile practices for mission and life-critical systems is dependent upon different aspects of the software system [5]. The next section presents the guidance Stage 2 provides on how to detect and discard inappropriate practices for mission and life-critical systems.

## 3.2. Stage 2: Discarding Inappropriate Practices

Stage 2 is responsible for providing an approach to determine the suitability of the individual agile practices relative to mission and life-critical systems. A practice is suitable if it is compatible with the following two aspects of mission and life-critical systems:

1. *System/Development Characteristics:* this refers to the set of attributes that describe the system in terms of both, its development *process* (e.g. long, expensive), and the actual *product* (e.g. large, complex).

2. *Desired System Qualities:* this refers to the desired benchmarks that describe the system's intended performance within the environment for which it is built (e.g. reliability) [6].

Stage 2 suggests four steps to identify inappropriate practices that need to be discarded. Figure 3 illustrates the four steps of Stage 2. The following two subsections elaborate on each of the steps needed to determine if agile practices conflict with system/development characteristics or preclude desirable system qualities of mission and life-critical systems.

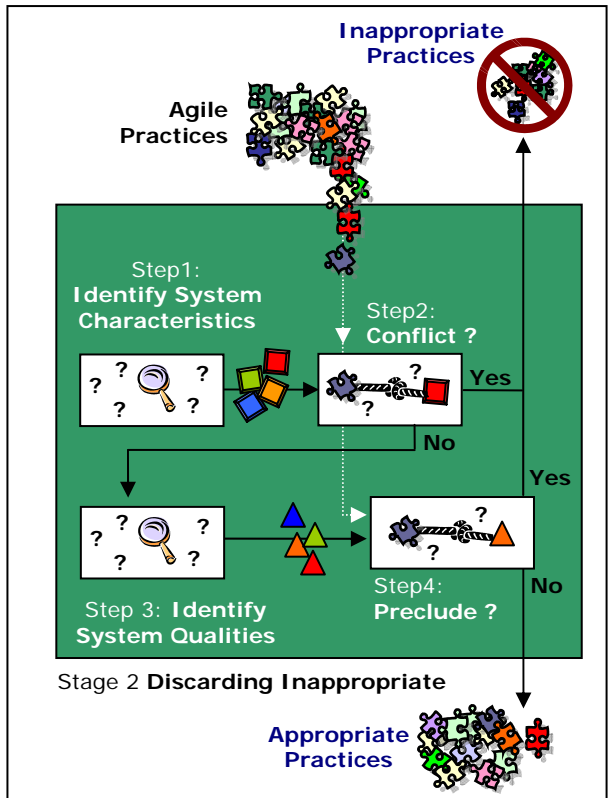

**Figure 3. The four steps of Stage 2**

**3.2.1. Synchronizing agile practices with system/development characteristics.** Two steps are carried out to ensure that an agile practice is compatible with the system/development characteristics.

*Step 1: Identify system/development characteristics.* These include (a) characteristics of the *software development process*, such as the cost and the duration of development, and (b) characteristics of the *software product*, such as the size of the application and the complexity of the product. Given that the type of software system usually determines these characteristics. Some of the typical system characteristics for mission or life-critical systems are:

- High product complexity
- Long development & maintenance periods
- Large applications .

*Step 2: Discover potential conflicts between agile practices and system characteristics.* Once the system characteristics are identified, the second step is to determine if an agile practice conflicts with the necessities of any of the characteristics. For example, the presence of long-life, persistent documentation for training and support is necessary for systems having a long development process. *Automated Unit Tests* (an agile practice) does not conflict with that need, but minimal documentation does. (as presented in Section 3.1.1.) Therefore, organizations should discard minimal documentation as an agile practice for adoption, because it conflicts with a system/development characteristic of mission and life-critical systems.

**3.2.2. Synchronizing agile practices with desirable systems qualities.** Two more steps are carried out to ensure that an agile practice does not prevent a system from realizing any of its desirable qualities.

*Step 3: Identify the desired system qualities.* Similar to the system characteristics, the type of software system determines the desired qualities of the system. For example, since the system is mission or life-critical, the system must exhibit a high degree of reliability. Generally, regulatory agencies require the development of mission and life-critical systems to exhibit certain enhanced system qualities [24], including:

- Safety
- Reliability
- Maintainability.

*Step 4: Discover agile practices that preclude desired system qualities.* If adopting an agile practice impedes the realization of any of these system qualities, then it is immediately determined to be unsuitable for mission and life-critical systems. For example, adopting evolutionary requirements reduces the capability for establishing high reliability (details presented in Section 3.1.2.) However, an agile practice like *Daily Standup Meetings* does not preclude the development of a system that embodies the desired qualities, and therefore is a suitable practice to adopt.

The four steps presented above serve as a guide for determining the suitability of individual agile practices with respect to mission and life-critical systems. An agile practice is considered unsuitable and must be discarded when it conflicts with integral system/development characteristics, or precludes the attainment of desirable qualities.

# 4. Determining the Suitability of Agile Practices for Organizations Developing Mission and Life-critical Systems

After following the steps outlined in Stage 2, the practices that remain are deemed suitable for mission and life-critical systems. However, this does not imply that the practice is *also* suitable for the organization developing that system. Agile practices suitable for mission and life-critical systems are selected for adoption only when they have been determined to be suitable for the *organization* developing the mission or life-critical system. This assertion motivates the third challenge related to introducing agile practices in mission and life-critical systems.

Section 4.1 elaborates on the importance of assessing the suitability of agile practices relative to the organization. Section 4.2 presents Stage 3: using organizational characteristics to determine whether or not an agile practice is suitable for adoption.

## 4.1. Benefits of assessing the suitability of agile practices for the organization.

Previous sections demonstrate that unsuitable practices can have detrimental impacts on the characteristics and desired qualities of mission and life-critical systems. Therefore spending time and effort to assess the suitability of each agile practice *relative to the software system* is justifiable. It might now seem like an unnecessary overhead to assess each agile practice *relative to the organization*. However, just the opposite is true. Investing time and effort in such an elaborate pre-adoption assessment of each agile practice is desirable, especially with mission and life-critical systems, because it significantly reduces the risks associated with the agile adoption process. When mismatches between the agile practice and the organization are identified before the adoption effort, the probability of failing during the adoption effort or introducing undesirable agile practices is reduced. Pre-adoption assessment helps to successfully introduce agile practices into traditional processes [12].

For example, assume *Pair Programming* is a suitable practice for developing mission and life-critical systems. If the organization developing the system has major collaboration issues, then adopting pair programming is likely to fail. This stems from the fact that pair programming works best within a collaborative environment.

To minimize the risks of failure when adopting agile practices in organizations that develop mission or life-critical systems, the suitability of each agile practice has to be examined relative to supporting (or inhibiting) characteristics of the development organization.

## 4.2. Stage 3: Determining the Right Practices to Adopt – An Organizational Perspective.

Stage 3 describes the process for determining the suitability of individual agile practices relative to characteristics of organizations that develop mission and life-critical systems. It uses two steps to determine the right practices to adopt:

1. *Identify candidate agile practices to adopt*
2. *Assess the readiness of the organization for each candidate practice.*

Figure 4 illustrates the above two steps defined in Stage 3; the following two subsections elaborate on them.

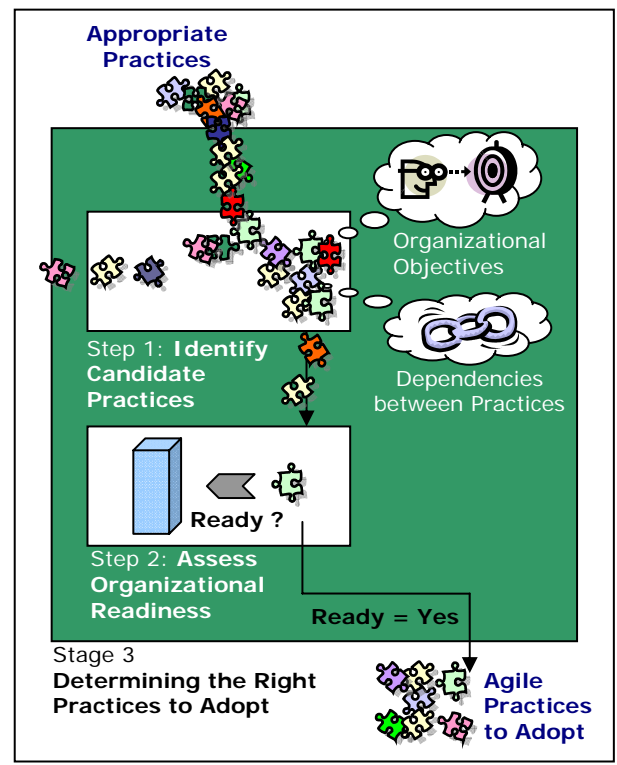

**Figure 4: The two steps of Stage 3**

### 4.2.1. Identify candidate agile practices to adopt.
There are numerous agile practices that are used for agile software development [30] [1]. Since assessing the suitability of all of the practices for organization is a costly (and unnecessary) approach, we must first choose a set of candidate practices from all available agile practices. As a result, the suitability of only these candidate practices is assessed with respect to the organization. Each candidate practice should be selected based on the following two factors [15] [12] :

- *Organizational Objectives.* To begin with, it is important to note that each agile practice is intended to help the organizations establish or realize certain business values or goals. For example, *Test Driven Development (TDD)* is an agile practice used to *increase the quality of the product delivered*. Pair programming is another practice that increases the quality of the product delivered, and helps address other goals, e.g. *lowering the employee turnover.* In view of this, organizations usually initiate agile adoption to achieve identified objectives or to promote new business values. For example, an organization that develops mission and life-critical systems would try to introduce agile practices to increase the quality of the product delivered. Therefore, organizations should select agile practices as candidates for agile adoption when they *assist in the realization of an objective for the agile adoption effort*.

- *Dependencies of the Practices.* Some practices depend on the presence of other practices during their adoption, while other practices can be adopted independently. Understanding the dependencies that exist between agile practices is an important factor when choosing candidate practices. Without considering these dependencies, an agile practice might be selected as a candidate practice for adoption while one of its prerequisites is not chosen. This type of mistake places the successful adoption of the practice at risk. Here are some examples of how practices are dependent on each other.

    a) Test driven development is dependent on the use of *Unit Testing* [26].

    b) Similarly, *Continuous Integration* is dependent on automated unit tests [10].

    c) Having *Self Organizing Teams* is dependent on having *Motivated and Empowered Individuals* [31].

By considering the organizational objectives and dependencies of the practices when selecting candidates for adoption, the chosen set of agile practices will be complementary and beneficial for the organization. But, how does one *really* determine if an agile practice is suitable for an organization?

**4.2.2. Assess readiness of organization for each candidate practice.** Before the organization invests time and money in adopting an agile practice it must be ready for it. The organization is ready for a practice once it possesses the characteristics necessary for the agile practice to function. The extent to which these organizational characteristics are present or absent is measured to determine the readiness of the organization for a practice.

For example, assume *Collaborative Planning* is a candidate practice for adoption. To assess the readiness of the organization to adopt collaborative planning, the following are some of the organizational characteristics that need to be present:

- Collaborative management style
- Management buy-in (to adopt the agile practice)
- Transparency of management
- Small power-distance in the organization
- Developers buy-in (to adopt the agile practice)

The assessor uses a set of indicators (questions) to measure the extent to which each of these organizational characteristics are present. Depending on the question, a manager, a developer, or the assessor himself or herself answers it. The Agile Adoption Framework provides a substantial set of indicators that are used to measure the organizational characteristics needed for approximately 30 different agile practices [29].

The lower the degree of presence of the organizational characteristics, the higher the risk is to successfully adopt the practice. Due to the significance of mission and life-critical systems, it is recommended that the organizational characteristics be fully present before the adoption process is attempted.

If the presence of all the characteristics needed by a particular agile practice is low, then the organization is not ready for that practice. If the organization cannot elevate these characteristics to adequate levels, the practice is considered unsuitable for the organization and its adoption is not recommended. On the other hand, the assessor selects an agile practice for adoption when the organizational characteristics needed for the practice are sufficiently present.

Therefore, to determine the right practices for an organization to adopt, the assessor chooses a set of practices based on the objectives of the organization. The assessor then determines whether or not the organization possesses the characteristics needed for each of the chosen practices.

## 5. Conclusions

While adopting all agile practices is not a viable option for mission and life-critical systems, certain agile practices are beneficial. The challenge is to identify the agile practices that improve the software development process, without causing detrimental impacts to any aspect of the software system. This paper has presented a 3-stage process for determining the applicability of adopting individual agile practices in mission and life-critical systems.

The first stage of the process provides assurance that the organization is capable of starting a process improvement effort involving the adoption of agile practices. Although not relevant to the actual agile practices that are to be adopted, this stage is necessary to identify the presence of specific *success factors* in the organization that establish its ability to pursue agile adoption. Since adopting an agile practice that is incompatible with mission and life-critical systems causes detrimental impacts to the software system, identifying the suitability of an agile practice with these systems becomes essential. Stage 2 of the process demonstrates why it is necessary to discard an agile practice if it conflicts with the *systems/development characteristics* or precludes *desirable system qualities*. The final stage in determining the applicability of an agile practice is to confirm that the *organization* possesses the characteristics necessary to implement the agile practice.

By following these three stages, organizations that develop mission and life-critical systems can take advantage of the benefits of agile practices. These three stages insure that the organization will select a set of agile practices that are (a) suitable for the organization, (b) achieve the organization's objectives, and (c) most importantly, do not conflict with any of the system's characteristics or impact any of the desired system qualities.